\documentclass[aps,prd,notitlepage,nofootinbib,preprintnumbers,superscriptaddress,10pt]{revtex4-2}
\usepackage{amsfonts,amssymb,amsmath,graphicx,color,bm}
\definecolor{indigo(dye)}{rgb}{0.0, 0.25, 0.42}
\usepackage[linktocpage=true]{hyperref}
\hypersetup{
colorlinks=true,
citecolor=blue, 
linkcolor=red, 
urlcolor=indigo(dye),
}
\usepackage{comment}

\begin{document}

\preprint{YITP-22-29, IPMU22-0018}

\title{
Static and spherically symmetric general relativity solutions\\
 in Minimal Theory of Bigravity
}

\author{Masato Minamitsuji}
\affiliation{Centro de Astrof\'{\i}sica e Gravita\c c\~ao  - CENTRA, Departamento de F\'{\i}sica, Instituto Superior T\'ecnico - IST, Universidade de Lisboa - UL, Av. Rovisco Pais 1, 1049-001 Lisboa, Portugal}

\author{Antonio De Felice}
\affiliation{Center for Gravitational Physics and Quantum Information, Yukawa Institute for Theoretical Physics, Kyoto University, 606-8502, Kyoto, Japan}

\author{Shinji Mukohyama}
\affiliation{Center for Gravitational Physics and Quantum Information, Yukawa Institute for Theoretical Physics, Kyoto University, 606-8502, Kyoto, Japan}
\affiliation{Kavli Institute for the Physics and Mathematics of the Universe (WPI), The University of Tokyo Institutes for Advanced Study, The University of Tokyo, Kashiwa, Chiba 277-8583, Japan}

\author{Michele Oliosi}
\affiliation{Center for Gravitational Physics and Quantum Information, Yukawa Institute for Theoretical Physics, Kyoto University, 606-8502, Kyoto, Japan}

\begin{abstract}
We investigate static and spherically symmetric solutions in the Minimal Theory of Bigravity (MTBG). First, we show that a pair of Schwarzschild-de Sitter spacetimes with different cosmological constants and black hole masses written in the spatially-flat Gullstrand-Painlev\'e (GP) coordinates is a solution in the self-accelerating branch of MTBG, while it cannot be a solution in the normal branch. We then illustrate how Schwarzschild-de Sitter solutions can become compatible with the normal branch when using different coordinates. We also confirm that the self-accelerating branch of MTBG admits static and spherically symmetric general relativity solutions with matter written in the spatially-flat coordinates, including neutron stars with arbitrary matter equations of state. Finally, we show that in the self-accelerating branch nontrivial solutions are given by the Schwarzschild-de Sitter metrics written in nonstandard coordinates. 
\end{abstract}
\maketitle

\section{Introduction}
\label{sec1}

While general relativity (GR) has passed all the experimental tests in
the weak-gravity field regime~\cite{Will:2014kxa}, with the dawn of
gravitational-wave astronomy and other experiments associated with
black holes (BHs) and compact objects a new frontier for testing GR
has
opened~\cite{Berti:2015itd,Berti:2018vdi,Berti:2018cxi,LIGOScientific:2021sio}.
A part of scalar-tensor theories which have passed Solar System tests
on GR could still have large deviations from GR in the vicinity of BHs
and compact stars.  Such a mechanism is known as spontaneous
scalarization~\cite{Damour:1993hw,Damour:1996ke,Harada:1997mr,Harada:1998ge,Novak:1998rk,Palenzuela:2013hsa,Sampson:2014qqa,Pani:2014jra,Silva:2014fca,Antoniou:2017acq,Silva:2017uqg,Doneva:2017bvd,Herdeiro:2018wub,Cunha:2019dwb,Cunha:2019dwb},
which is triggered by a tachyonic instability on a BH or compact star
background in GR.  A key assumption for successful spontaneous
scalarization is the existence of {\it GR solutions}, namely, the
solutions that share the same metric and matter profiles with those in
GR.
While certain non-GR metric gravitational theories could allow GR
solutions \cite{Psaltis:2007cw,Motohashi:2018wdq}, perturbations on
top of them can behave very differently from those in GR
\cite{Barausse:2008xv}.

There were several attempts to extend the idea of spontaneous
scalarization to other field species
\cite{Ramazanoglu:2017xbl,Ramazanoglu:2019gbz,Annulli:2019fzq,Ramazanoglu:2019jrr,Kase:2020yhw,Minamitsuji:2020pak,Minamitsuji:2020hpl,Demirboga:2021nrc},
for instance a vector field.  However, it has been argued that
spontaneous vectorization
\cite{Ramazanoglu:2017xbl,Ramazanoglu:2019gbz,Annulli:2019fzq,Ramazanoglu:2019jrr,Kase:2020yhw}
can be realized from selected initial conditions, rather than from the
tachyonic instabilities of GR solutions. Since the branch of
vectorized solutions is disconnected from the GR branch, spontaneous
vectorization would not proceed as a continuous evolution from a GR
solution.  Indeed, in GR solutions the scalar mode suffers from ghost
instabilities, rather than tachyonic ones
\cite{Silva:2021jya,Minamitsuji:2020pak,Demirboga:2021nrc}.  A similar
problem has also been pointed out for the models of spontaneous
spinorization \cite{Silva:2021jya}.  One important lesson from such
models is that for a successful higher-spin extension of spontaneous
scalarization the gravitational theory should, on top of having GR
solutions, not contain any extra scalar modes.

We expect that
{\it spontaneous tensorization}
which is analogous to spontaneous scalarization in the spin-2 sector
could occur in the fiducial sector of a bigravity theory,
if the theory does not contain any extra
scalar and vector degrees of freedom (DOFs)
besides the two metric sectors.
The de Rham-Gabadadze-Tolley (dRGT) massive gravity~\cite{deRham:2010kj}
which was the first nonlinear massive gravity theory
free from the Boulware–Deser (BD) ghost \cite{Boulware:1972yco}
has been extended to bigravity by Hassan and Rosen (HR) \cite{Hassan:2011zd}
by promoting the metric of the fiducial sector to a dynamical field. 
In HR bigravity, however, the BD ghost can be generically revived
when the matter is coupled to both the physical and fiducial metrics
\cite{Yamashita:2014fga,deRham:2014fha}.
The Minimal Theory of Massive Gravity (MTMG)
\cite{DeFelice:2015hla,DeFelice:2015moy} is an extension of dRGT
massive gravity with only two tensorial DOFs as in GR, instead of 5
DOFs in dRGT. In the original formulation of MTMG, the
four-dimensional diffeomorphism invariance is completely broken as the
unitary gauge for both time and space directions is chosen.  Cosmology
in MTMG has been studied in Refs.\
\cite{DeFelice:2015moy,DeFelice:2016ufg,Hagala:2020eax,DeFelice:2021trp},
where cosmological solutions in MTMG have been classified into the
normal and self-accelerating branches.  BH and stellar solutions in
MTMG have been investigated in Ref.\ \cite{DeFelice:2018vza}, where it
has been shown that any GR solution which can be written in terms of
the spatially-flat Gullstrand–Painlev\'e (GP) coordinates can be a
solution in the self-accelerating branch of MTMG. Up to now, still spherically symmetric solutions of
MTMG in the normal branch are not known,
as in general
the normal branch is harder to study than the self-accelerating branch\footnote{The property that normal branch
  solutions are more difficult to be found, as we will see, is also
  shared by the theory we will discuss here.}.

The ideas behind MTMG have been applied to the case of bigravity,
which led to the Minimal Theory of Bigravity (MTBG)
\cite{DeFelice:2020ecp}, where this time the joint four-dimensional
diffeomorphism invariance is broken down to the three-dimensional one.
While MTBG shares, by construction, the same background cosmological dynamics with HR
bigravity, the number of propagating DOFs are down to four, namely,
two tensorial DOFs in the physical metric and the other two in the
fiducial metric, at least in the absence of matter. The absence of
the extra scalar and vector DOFs in MTBG means the absence of ghost/gradient
instabilities associated with them, which is one of the desired features
to successfully realize spontaneous tensorization, and motivates us to
study BHs and stars in MTBG.  In order to see whether spontaneous
tensorization can be successfully realized in MTBG, in this work we
are going to clarify the existence of static and spherically symmetric
GR solutions without and with matter in MTBG. Of course,
even if spontaneous tensorization cannot be realized,
these solutions would also remain
interesting per se,
as one can hope to test MTBG in the strong gravity regime by
investigating perturbations around them.

Therefore, we will study static and spherically symmetric solutions in
MTBG with and without matter.  We will clarify the conditions under
which static and spherically symmetric GR solutions written in the
spatially-flat coordinates are also solutions in MTBG, and explicitly
show that Schwarzschild-de Sitter solutions written in the
spatially-flat GP coordinates can also be solutions in the
self-accelerating branch of MTBG.  We will also show that in the
self-accelerating branch of MTBG with two matter sectors coupled to
the two metrics separately the static and spherically symmetric
physical and fiducial metrics written in the spatially-flat
coordinates satisfy the Einstein equations with matter in GR, namely,
GR solutions with two individual matter sectors can also be solutions
in MTBG.  Finally we will show how Schwarzschild-de Sitter solutions,
when written in other non-spatially-flat coordinates, may become
solutions in the normal branch as well, provided that the two metrics
are parallel to each other. This last condition, which does not need
to hold in the self-accelerating branch, shows indeed that far fewer
solutions are known for MTBG in the normal branch.

On the other hand, in order to investigate whether MTBG admits
nontrivial static and spherically symmetric solutions besides GR
solutions, we will construct static and spherically symmetric vacuum
solutions perturbatively in the small mass limit of MTBG.  We will
regard the graviton mass squared as the expansion parameter, and
expand the metric variables and the Lagrange multipliers.  We will
show that in the self-accelerating branch the Schwarzschild-de Sitter
metrics written in nonstandard coordinates can be obtained
perturbatively, while in the normal branch these metrics are not
compatible with the massless limit of MTBG.  We will also show that
the results in the perturbative analysis can be naturally extended to
the fully nonperturbative analysis, and the Schwarzschild-de Sitter
solutions written in the nonstandard coordinates satisfying certain
conditions can be solutions in the self-accelerating branch of MTBG.

The structure of this paper is as follows: In Sec.\ \ref{sec2}, we
review the formulation of MTBG theory.  In Sec.\ \ref{sec3}, we derive
the conditions under which static and spherically symmetric vacuum
solutions, i.e., the Schwarzschild-de Sitter solutions written in the
spatially-flat GP coordinates are also solutions in MTBG.  In
Sec.\ \ref{sec4}, we consider the case where two individual matter
sectors are coupled to the two metrics separately, and show that the
static and spherically symmetric physical and fiducial metrics written
in the spatially-flat coordinates satisfy the Einstein equations with
matter in GR, and hence GR solutions with matter coupled to each
sector separately can also be solutions in MTBG.  In Sec.\ \ref{sec5},
we investigate static and spherically symmetric vacuum solutions with
the perturbative and nonperturbative approaches, and show the
existence of the Schwarzschild-de Sitter solutions written in
nonstandard coordinates.  Finally, Sec.\ \ref{sec6} is devoted to a brief
summary and conclusion.

\section{Minimal theory of bigravity}
\label{sec2}

We consider bigravity composed of the physical and fiducial metrics,
$g_{\mu\nu}$ and $f_{\mu\nu}$, respectively.
We introduce the Arnowitt-Deser-Misner (ADM) decomposition
of $g_{\mu\nu}$ and $f_{\mu\nu}$, respectively, 
\begin{eqnarray}
&&
\label{adm}
g_{\mu\nu}dx^\mu dx^\nu
=-N^2dt^2
+\gamma_{ij}
\left(dx^i+N^i dt\right)
\left(dx^j+N^j dt\right),
\nonumber\\
&&
f_{\mu\nu}dx^\mu dx^\nu
=-M^2dt^2
+\phi_{ij}
\left(dx^i+M^i dt\right)
\left(dx^j+M^j dt\right),
\end{eqnarray}
where 
$t$ and $x^i$ ($i=1,2,3$) are the temporal and spatial coordinates, 
$(N,N^i,\gamma_{ij})$ 
and
$(M,M^i,\phi_{ij})$ 
are the sets of 
the lapse function, shift vector, and spatial metric 
on the constant $t$ hypersurfaces
for the metrics $g_{\mu\nu}$ and $f_{\mu\nu}$,
respectively.
The extrinsic curvature tensors on constant $t$ hypersurfaces
are given by 
\begin{eqnarray}
\label{extrinsic}
K_{ij}
:=
\frac{1}{2N}
\left(
\partial_t \gamma_{ij}
-D_i N_j
-D_j N_i
\right),
\quad 
\Phi_{ij}
:=
\frac{1}{2M}
\left(
\partial_t \phi_{ij}
-{\tilde D}_i M_j
-{\tilde D}_j M_i
\right),
\end{eqnarray}
where $D_i$ and ${\tilde D}_i$ are covariant derivatives for $\gamma_{ij}$ and $\phi_{ij}$,
respectively.

The action of MTBG is then given by 
\begin{eqnarray}
\label{action}
S
&=&
\frac{1}{2\kappa^2}
\int d^4x {\cal L},
\nonumber\\
{\cal L}
&:=&
{\cal L}_g
\left[N,N^i,\gamma_{ij}; M,M^i,\phi_{ij};
\lambda, {\bar \lambda}, \lambda^i
\right]
+
{\cal L}_m
\left[
N,N^i,\gamma_{ij};
M,M^i,\phi_{ij};
\Psi
\right],
\end{eqnarray}
where 
${\cal L}_g$ and ${\cal L}_m$
represent the gravitational and matter parts of the Lagrangian,
respectively.
$\lambda$, ${\bar \lambda}$, and $\lambda^i$
are the Lagrange multipliers,
and 
$\Psi$ represents matter.
The gravitational Lagrangian ${\cal L}_g$ of MTBG in the unitary gauge
is further decomposed into
the {\it precursor} and {\it constraint} parts as 
\begin{eqnarray}
{\cal L}_g
&:=&
{\cal L}_{\rm pre}
\left[
N,N^i,\gamma_{ij};
M,M^i,\phi_{ij}
\right]
+
{\cal L}_{\rm con}
\left[N,N^i,\gamma_{ij};
M,M^i,\phi_{ij};
\lambda, {\bar \lambda}, \lambda^i
\right],
\end{eqnarray}
with
\begin{eqnarray}
\label{lag2}
{\cal L}_{\rm pre}
&:=&
\sqrt{-g}R[g]
+
\tilde{\alpha}^2
\sqrt{-f} R[f]
-
m^2
\left(
N\sqrt{\gamma}{\cal H}_0
+
M\sqrt{\phi}\tilde{\cal H}_0
\right),
\nonumber
\\
{\cal L}_{\rm con}
&:=&
\sqrt{\gamma}
\alpha_{1\gamma}
\left(
\lambda
+
\Delta_\gamma
{\bar \lambda}
\right)
+
\sqrt{\phi}
\alpha_{1\phi}
\left(
\lambda
-
\Delta_\phi
{\bar \lambda}
\right)
+
\sqrt{\gamma}
\alpha_{2\gamma}
\left(
\lambda
+
\Delta_\gamma
 {\bar\lambda}
\right)^2
+
\sqrt{\phi}
\alpha_{2\phi}
\left(
\lambda
-\Delta_\phi {\bar\lambda}
\right)^2
\nonumber\\
&-&
m^2
\left[
\sqrt{\gamma}
  U^i{}_k D_i \lambda^k
-\beta 
\sqrt{\phi}
{\tilde U}_k{}^i {\tilde D}_i \lambda^k
\right],
\end{eqnarray}
and 
\begin{eqnarray}
\alpha_{1\gamma}
&:=&
-m^2
 U^p{}_q K^q{}_p,
\quad 
\alpha_{1\phi}
:=
m^2 {\tilde U}^p{}_q
    \Phi^q{}_p,
\nonumber\\
\alpha_{2\gamma}
&:=&
-\frac{m^4 }{4N}
\left(
U^p{}_q-\frac{1}{2}U^k{}_k \delta^p{}_q
\right)
U^q{}_p,
\quad 
\alpha_{2\phi}
:=
-\frac{m^4 }{4M\tilde{\alpha}^2}
\left(
{\tilde U}_q{}^p
-\frac{1}{2}{\tilde U}_k{}^k \delta_q{}^p
\right)
{\tilde U}_p{}^q,
\end{eqnarray}
where the constant $\tilde{\alpha}$ is the ratio of the two
gravitational constants, $m$ is a parameter with dimensions of mass
which can be related with the graviton mass, $\beta$ is a constant,
$\gamma:={\rm det} (\gamma_{ij})$ and $\phi:={\rm det} (\phi_{ij})$
are the determinants of the three-dimensional spatial metrics
$\gamma_{ij}$ and $\phi_{ij}$ respectively. Furthermore, ${\cal H}_0$
and $\tilde{\cal H}_0$ are defined by
$ {\cal H}_0 := \sum_{n=0}^3 c_{4-n} e_n( {\cal K})$ and
$ \tilde{\cal H}_0 := \sum_{n=0}^3 c_{n} e_n( \tilde{\cal K})$ with
\begin{eqnarray}
e_0({\cal K})
=1,
\quad
e_1({\cal K})
=
\left[
{\cal K}
\right],
\quad 
e_{2} ({\cal K})
=
\frac{1}{2}
\left(
\left[
{\cal K}
\right]^2
-
\left[
{\cal K}^2
\right]
\right),
\quad
e_{3} ({\cal K})
=
{\rm det} 
({\cal K}),
\end{eqnarray}
and similarly for $e_n (\tilde{\cal K})$
with ${\cal K}^i{}_k$ and $\tilde{\cal K}_k{}^i$
characterized by
\begin{eqnarray}
\label{calk}
{\cal K}^i{}_k
{\cal K}^k{}_j
=
{\gamma}^{ik}
\phi_{kj},
\quad 
{\cal {\tilde K}}_j{}^k
{\cal {\tilde K}}_k{}^i
=
\gamma_{jk}
{\phi}^{ki},
\end{eqnarray}
$\Delta_\gamma :=\gamma^{ij} D_i D_j$
and 
$\Delta_\phi :=\phi^{ij} {\tilde D}_i {\tilde D}_j$
are
the Laplacian operators,
the spatial tensors
$U^i{}_j$ and ${\tilde U}{}_j{}^i$ are defined by
\begin{eqnarray}
U^i{}_j
&:=&
\frac{1}{2}
\sum_{n=1}^3 c_{4-n}
\left(
U_{(n)}{}^i{}_j
+\gamma^{ik}\gamma_{j\ell}
U_{(n)}{}^\ell{}_k
\right),
\nonumber\\
{\tilde U}_j{}^i
&:=&
\frac{1}{2}
\sum_{n=1}^3 c_{n}
\left(
{\tilde U}_{(n)j}{}^i
+\phi^{ik}\phi_{j\ell}
{\tilde U}_{(n)k}{}^\ell
\right),
\end{eqnarray}
with
$U_{(n)}{}^i{}_k
:=
\frac{\partial e_n ({\cal K})}
        {\partial {\cal K}^k{}_i}$
and  
${\tilde U}_{(n)k}{}^i
:=
\frac{\partial e_n ({\cal {\tilde K}})}
        {\partial \tilde{\cal K}^k{}_i}$,
and $c_j$ ($j=0,1,2,3,4$) being constants.

Variation of the action \eqref{action}
with respect to $N$, $N^i$, $\gamma_{ij}$, $M$, $M^i$, and $\phi_{ij}$
provides the gravitational equations of motion,
and 
variation with respect to $\Psi$
provides the matter equation of motion,
which we do not show explicitly.
Finally,
variation with respect to 
the Lagrange multipliers
$\lambda$, ${\bar \lambda}$, and $\lambda^i$
gives the constraint equations
\begin{eqnarray}
\label{cons2}
&&
  \sqrt{\gamma}\alpha_{1\gamma}
+\sqrt{\phi}\alpha_{1\phi}
+
2\sqrt{\gamma} \alpha_{2\gamma}
\left(
\lambda+
\Delta_\gamma
{\bar \lambda}
\right)
+
2\sqrt{\phi} \alpha_{2\phi}
\left(
\lambda
-
\Delta_\phi
{\bar \lambda}
\right)=0,
\\
&&
\label{cons3}
\sqrt{\gamma}
\Delta_{\gamma} \alpha_{1\gamma}
-
\sqrt{\phi}
\Delta_{\phi}\alpha_{1\phi}
+
2\sqrt{\gamma} 
\Delta_\gamma
\left[
\alpha_{2\gamma}
\left(
\lambda
+
\Delta_\gamma{\bar \lambda}
\right)
\right]
-
2\sqrt{\phi} 
\Delta_\phi
\left[
\alpha_{2\phi}
\left(
\lambda
-
\Delta_\phi
{\bar \lambda}
\right)
\right]
=0,
\\
\label{const}
&&
\sqrt{\gamma} D_p U^p{}_q
-\beta
\sqrt{\phi} {\tilde D}_p {\tilde U}_q{}^p
=0.
\end{eqnarray}
It should be noted that adding constraints to a theory in general shrinks the space for the allowed solutions. As for MTBG, these same constraints are, however, necessary in order to remove the unwanted unstable degrees of freedom, while keeping the tensor modes for both metrics propagating on any background solutions of the theory. On a homogeneous and isotropic background, it was shown that in general nontrivial solutions do exist. However, on a different background, things could go differently, and we would like to investigate in the following the presence of solutions in spherically symmetric configurations.

\section{Schwarzschild-de Sitter solutions in MTBG}
\label{sec3}

First, we consider the vacuum case 
by setting the matter action to be zero ${\cal L}_m=0$ in the action \eqref{action}.
We derive the conditions 
under which static and spherically symmetric solutions
written in the spatially-flat coordinates
in vacuum GR,
namely,
the Schwarzschild-de Sitter solutions in the GP coordinates, 
are also the solutions in MTBG.
We consider the static and spherically symmetric physical and fiducial metrics
written in the following coordinates
\begin{eqnarray}
\label{generalfg}
g_{\mu\nu}
dx^\mu dx^\nu
&=&
-A_0(r)dt^2
+A_1(r)
\left(dr+N^r(r) dt\right)^2
+A_2(r)r^2 
\left(
 d\theta^2
+\sin^2\theta d\varphi^2
\right),
\nonumber\\
f_{\mu\nu}
dx^\mu dx^\nu
&=&
C_0^2
\left[
- A_{0f}(r)dt^2
+A_{1f}(r)\left(dr+N^r_f(r) dt\right)^2
+A_{2f}(r)
r^2 
\left(
 d\theta^2
+\sin^2\theta d\varphi^2
\right)
\right],
\end{eqnarray}
where $C_0>0$ is constant, 
$r$ and $(\theta,\varphi)$ represent the radial and angular coordinates, 
and 
$A_0(r)$, $A_1(r)$, $A_2(r)$,
$A_{0f}(r)$, $A_{1f}(r)$, and $A_{2f}(r)$
are functions of $r$.

\subsection{Self-accelerating branch in GP coordinates}

The Schwarzschild-de Sitter metric in the spatially-flat GP coordinates
is expressed as 
\begin{eqnarray}
\label{gp2}
&&
A_0(r)=A_1(r)=A_2(r)=1,
\quad 
N^r(r)
=
\pm
\sqrt{
\frac{2M}{r}
+\frac{\Lambda}{3}r^2},
\nonumber\\
&&
A_{0f}(r)=A_{1f}(r)=A_{2f}(r)=1,
\quad 
N^r_f(r)
=
\pm
\sqrt{
\frac{2M_f}{r}
+\frac{C_0^2\Lambda_f}{3}r^2},
\end{eqnarray}
where $M$ and $M_f$ represent the mass parameters,
and  
$\Lambda$ and $\Lambda_f$ are effective cosmological constants
in the physical and fiducial sectors, respectively.
In general, 
the masses and effective cosmological constants in the two sectors
may be different,  $M\neq M_f$ and $\Lambda\neq \Lambda_f$.

First,
focusing on the precursor part of the vacuum MTBG ${\cal L}_{\rm pre}$ in Eq.\ \eqref{lag2},
the conditions under which the Schwarzschild-de Sitter solutions exist are given by 
\begin{eqnarray}
\label{cond}
2\Lambda
=
\left(
   C_0^3c_1
+ 3C_0^2 c_2
+ 3C_0 c_3
+ c_4
\right)
m^2,
\quad 
2\tilde{\alpha}^2\Lambda_f
=
\left(
    c_0
+ 3C_0^{-1} c_1
+  3C_0^{-2} c_2
+ C_0^{-3} c_3
\right)
m^2.
\end{eqnarray}
We then consider the full vacuum MTBG ${\cal L}_g$
including the constraint part of the Lagrangian ${\cal L}_{\rm con}$ given by Eq.\ \eqref{lag2}.
Because of the symmetries in the
physical and fiducial sectors,
from the beginning we may choose
\begin{eqnarray}
\label{lambda_a_vanish}
\lambda^\theta(r)=\lambda^\varphi(r)=0,
\end{eqnarray}
while still assume that $\lambda^r(r)$ is a priori a nontrivial function of $r$.
For the Schwarzschild-de Sitter solutions
written in the spatially-flat GP coordinates \eqref{gp2},
we find that the constraint equation \eqref{const} is trivially satisfied.
The coefficients in the remaining constraint equations \eqref{cons2} and \eqref{cons3}
are given by 
\begin{eqnarray}
&&
\alpha_{1\gamma}
=\frac{(C_0^2 c_1+2C_0 c_2+ c_3)m^2(3M+r^3\Lambda)}
         {r^2\sqrt{\frac{2M}{r}+ \frac{\Lambda r^2}{3}}},
\quad 
\alpha_{1\phi}
=
-\frac{(C_0^2 c_1+2C_0 c_2+ c_3)m^2(3M_f+C_0^2 r^3\Lambda_f)}
         {r^2C_0^3\sqrt{\frac{2M_f}{r}+ \frac{\Lambda_f  C_0^2 r^2}{3}}},
\nonumber\\
&&
\alpha_{2\gamma}
=
\frac{3(C_0^2 c_1+2C_0 c_2+ c_3)^2m^4}{8},
\quad 
\alpha_{2\phi}
=
\frac{3(C_0^2 c_1+2C_0 c_2+ c_3)^2m^4}{8C_0^5\tilde{\alpha}^2},
\end{eqnarray}
and then Eqs.~\eqref{cons2} and \eqref{cons3} have the structure
\begin{eqnarray}
\label{const_structure}
&&
\left(
C_0^2 c_1+2C_0c_2+c_3
\right)
{\cal F} \left(
\lambda(r), 
{\bar \lambda}'(r),
{\bar \lambda}''(r)
\right)
=0,
\nonumber\\
&&
\left(
C_0^2 c_1+2C_0c_2+c_3
\right)
{\cal G} \left(
\lambda(r),
\lambda'(r), 
\lambda''(r),
{\bar \lambda}''(r),
{\bar \lambda}^{(3)}(r),
{\bar \lambda}^{(4)}(r)
\right)
=0,
\end{eqnarray}
where ${\cal F}$ and ${\cal G}$ are linear combinations of the variables in the argument.
From Eq.\ \eqref{const_structure}, 
we see that there are two possible branches,
i.e., the {\it self-accelerating} and {\it normal} branches.

The self-accelerating branch is given by 
\begin{eqnarray}
\label{additional}
C_0^2 c_1+2C_0c_2+c_3=0.
\end{eqnarray}
Substituting
the Schwarzschild-de Sitter metrics
written in the spatially-flat coordinates \eqref{gp2}
with the conditions 
\eqref{cond} and \eqref{additional} into the gravitational equations of motion of MTBG,
we find
\begin{eqnarray}
\label{laplace_conditions}
\lambda(r)=0,
\quad 
{\bar \lambda}''(r)
+\frac{2}{r}
{\bar \lambda}'(r)
=0,
\quad 
\lambda^r(r)
=0,
\end{eqnarray}
which allows us to choose
\begin{eqnarray}
\label{all_lambda_vanish}
\lambda(r)=\lambda^r(r)=0,
\qquad 
{\bar \lambda}(r)
=
d_0
+
\frac{d_1}{r},
\end{eqnarray}
with $d_0$ and $d_1$ being constants, as a solution.
Thus,  Eqs.\ \eqref{cond} and \eqref{additional}
provide the condition for the Schwarzschild-de Sitter metrics
written in the spatially-flat coordinates
 to be solutions in MTBG.
We note that the choice \eqref{additional} coincides with
the self-accelerating branch in cosmology \cite{DeFelice:2020ecp}.

\subsection{Normal branch in GP coordinates}

If a consistent solution of
\begin{eqnarray}
\label{fgzero}
{\cal F} \bigl(
\lambda(r), 
{\bar \lambda}'(r),
{\bar \lambda}''(r)
\bigr)
=0,
\quad
{\cal G} \bigl(
\lambda(r),
\lambda'(r), 
\lambda''(r),
{\bar \lambda}''(r),
{\bar \lambda}^{(3)}(r),
{\bar \lambda}^{(4)}(r)
\bigr)
=0,
\end{eqnarray}
exists for $\lambda(r)$ and ${\bar \lambda}(r)$,
this should correspond to the {\it normal branch}.
Within the Schwarzschild-de Sitter ansatz \eqref{generalfg} and \eqref{gp2},
for simplicity we focus on the asymptotically flat
Schwarzschild spacetimes with $\Lambda=\Lambda_f=0$.
The condition ${\cal F}=0$ in Eq.~\eqref{fgzero} then reduces to
\begin{eqnarray}
\label{lambda_sol}
\lambda(r)
&=&
\frac{1}{C_0^2 r^2 (1+C_0^2{\tilde \alpha}^2)}
\left[
\frac{2\sqrt{2}C_0^4{\tilde\alpha}^2}
       {m^2(C_0^2c_1+2C_0 c_2+c_3)}
\left(
\sqrt{M_fr}
-\sqrt{M r}
\right)
-r 
\left(-1+C_0^4{\tilde\alpha}^2\right)
\left(
2{\bar \lambda}'(r)
+r {\bar\lambda}''(r)
\right)
\right].
\end{eqnarray}
Then, imposing ${\cal G}=0$ in \eqref{fgzero} yields
\begin{eqnarray}
\label{blambda}
{\bar \lambda} (r)
=
-\frac{8C_0^2 \sqrt{2r}\left(\sqrt{M}+C_0^2{\tilde\alpha}^2 \sqrt{M_f}\right)}
          {3 (1+C_0^2) (C_0^2 c_1+2C_0 c_2+c_3)m^2}
+\frac{q_1}{r}
+q_2
+q_3 r
+q_4 r^2,
\end{eqnarray}
where $q_1$, $q_2$, $q_3$, and $q_4$
are integration constants.
We note that
the integration constants $q_1$ and $q_2$ are associated with
the solutions in the Laplace equation in the three-dimensional flat space
and do not physically contribute to the solution of  $\lambda(r)$, Eq.~\eqref{lambda_sol}.
On the other hand, 
the terms of $q_3$ and $q_4$ physically contribute to  Eq.~\eqref{lambda_sol}.
However, the Euler-Lagrange equations for $N^r(r)$ and $N_f^r(r)$
provide, respectively,
\begin{eqnarray}
&&
-3\sqrt{\frac{2M}{r}}
+\frac{2 (1+C_0^2) (c_3+2C_0c_2+c_1C_0^2)m^2 q_3}
         {C_0^2 (1+{\tilde\alpha}^2C_0^2)}=0,
\nonumber\\
&&
-3\sqrt{\frac{2M_f}{r}}
+\frac{2 (1+C_0^2) (c_3+2C_0c_2+c_1C_0^2)m^2 q_3}
         {C_0^2 (1+{\tilde\alpha}^2C_0^2)}=0,
\end{eqnarray}
whose combination provides $M=M_f$.
Assuming that $c_3+2C_0c_2+c_1C_0^2\neq 0$
for the non-self-accelerating branch,
they cannot be satisfied
unless $q_3=0$ and $M=M_f=0$.
Adding nonzero $\Lambda$ and $\Lambda_f$
does not change the results.
This means that in the normal branch,
Schwarzschild BHs
written in the spatially-flat GP coordinates
cannot be embedded into the vacuum spacetimes.

The same conclusion could be derived
for a regular matter distribution.
Thus,
we conclude that 
the normal branch of MTBG
cannot accommodate static and spherically symmetric GR solutions written in the spatially-flat GP coordinates,
irrespective of the presence of the matter sector
(see Appendix~\ref{app:nogo-normalbranch} for a similar result in the massless limit of the normal branch of MTBG).

\subsection{Normal branch in slicing with $D_iD^iK=0$}

The fact that the GP choice for the slicing of the metric does not satisfy the equations of motion does not necessarily mean that there are no solutions in the normal branch. 
In fact, even if the constraints imposed on MTBG
do not allow the GP slicing in the normal branch, in principle
there could be other slicings which could instead lead to
some nontrivial solutions for the equations of motion. Indeed, in MTBG, 
since the four-dimensional diffeomorphism invariance is broken, 
different time-slicings in general lead to different physical configurations.
Then, at least in principle, one is supposed to investigate the most general ansatz
for the differential equations, i.e.\ to consider all the different time-slicings 
which are compatible with a given background. 
And this attempt should be done, in principle, also 
for general nonzero values of $\lambda$, $\bar\lambda$ and $\lambda^r$.

Although this kind of general configurations would manifestly
show the whole background possibilities for the theory
\footnote{For MTBG, and for other theories which break four dimensional diffeomorphism,
a Birkhoff theorem does not hold in general, but one can still try to find out general solutions
to the equations of motion.},
in practice to do so turns out to be an analytically formidable problem. 
So, let us try to find instead a class of time-slicings
in the normal branch which can shed some light on the space of the solutions. 
In the remaining part of this section, 
we will also try to be as independent as possible from a given choice of the background, but only assume that the chosen slicing is compatible with the ADM splitting. Although at the beginning we will let the slicing to be general, still we aim to find a particular class of solutions (how particular will be discussed later on).

Hence, we focus our attention to a vacuum configuration solution which satisfies the following ansatz 
\begin{alignat}{6}
  \phi_{ij}&=C_0^{2}\,\gamma_{ij}\,,\qquad& M& =C_0\,N\,,\qquad & M^i&=N^i\,, \nonumber  \\
  \lambda&=0\,,\qquad & \bar\lambda&=\textrm{constant}\,,\qquad & 
 \lambda^i&=0\,,\label{eq:norm_ansatz}
\end{alignat}
where $C_0$ is again a numerical constant. Then this ansatz leads to the following relations on a general background
\begin{alignat}{8}
\tilde{D}_{j}A^{i}&=D_{j}A^{i}\,,\qquad&\tilde{D}_{j}B_{i}&=D_{j}B_{i}\,,\qquad &\tilde{D}_k\gamma_{ij}&=0\,,& M_i&=C_0^2\,N_i\,,\\
{}^{(3)}\tilde{R}_{ij}&={}^{(3)}R_{ij}\,,\qquad& {}^{(3)}\tilde{R}&=\frac{1}{C_0^{2}}\,{}^{(3)}R\,,\qquad& \phi^{ij}&=\frac{1}{C_0^{2}}\,\gamma^{ij}\,,\qquad & \phi&=C_0^{6}\,\gamma\,,
\end{alignat}
where $A^i$ ($B_i$) is a general three dimensional vector (covector), and evidently ${}^{(3)}R_{ij}$ and ${}^{(3)}R$ represent the three dimensional Ricci tensor and scalar for the metric $\gamma_{ij}$ (leaving a clear interpretation for ${}^{(3)}\tilde{R}_{ij}$ and ${}^{(3)}\tilde{R}$). On using the properties of the chosen ansatz, we find
\begin{equation}
\tilde{\mathcal{K}}^{i}{}_{k}\tilde{\mathcal{K}}^{k}{}_{j}=\gamma^{ik}\phi_{kj}=C_0^{2}\,\delta^{i}{}_{j}\,,
\end{equation}
so that
\begin{equation}
\tilde{\mathcal{K}}^{i}{}_{j}=C_0\,\delta^{i}{}_{j}\,,\qquad\mathcal{K}^{i}{}_{j}=\frac{1}{C_0}\,\delta^{i}{}_{j}\,,
\end{equation}
since $\mathcal{K}^i{}_j$ must be the inverse of $\tilde{\mathcal{K}}^i{}_j$. These expressions lead to the following relation for the extrinsic curvature for the two three dimensional metrics
\begin{equation}
\Phi_{ij}=C_0\,K_{ij}\,.
\end{equation}
We note that, so far, the value of $C_0$ is still a free numerical parameter.

In order to simplify and to solve all the constraints of MTBG in the normal branch, we impose the following condition on top of the ansatz (\ref{eq:norm_ansatz}). 
\begin{equation}
D_{i}D^{i}K=0\,.\label{VCDM_slicing}
\end{equation}
 At this point, one is left to solve the equations of motion for the metric.
Hence, on considering the equation of motion for the lapse $N$, in MTBG, we find that
\begin{equation}
K_{ij}K^{ij}-K^{2}-R+2\Lambda_{\rm eff}=0\,, 
\end{equation}
where we identify
\begin{equation}
\Lambda_{{\rm eff}}=\frac{1}{2}m^{2}\left(c_{1}C_0^{3}+3c_{2}C_0^{2}+3c_{3}C_0+c_{4}\right)\,.
\end{equation}
Here we remind the reader that the term $c_4$ is a pure cosmological constant 
in the physical sector, 
whereas $c_0$ is a pure cosmological constant in the fiducial sector.
As for the equation of motion for the lapse function in the fiducial sector $M$,
we have that, on combining it with the previous 
equation of motion for the lapse function in the physical sector $N$, 
the following condition must hold, namely
\begin{equation}
c_{0}C_0^{3}+c_{1}C_0^{2}\left(3-\alpha^{2}C_0^{2}\right)+3c_{2}C_0\left(1-\alpha^{2}C_0^{2}\right)+c_{3}\left(1-3\alpha^{2}C_0^{2}\right)-\alpha^{2}c_{4}C_0=0\,.\label{eq:defC}
\end{equation}
This constraint is \textit{not} to be understood as a fine-tuning condition for the fiducial cosmological constant $c_0$ (or vice versa for the physical cosmological constant $c_4$), rather it is the equation which determines $C_0$ for the solution. So at this level, whatever the cosmological constant are in the two metric frames, a solution will in general exist, provided a real $C_0$ satisfies Eq.\ \eqref{eq:defC}.

Now, let us consider the remaining nontrivial Hamilton equations of motion for the two metrics, namely
\begin{eqnarray}
\dot{\pi}^{ij} & = & \{\pi^{ij},H\}\,,\\
\dot{\sigma}^{ij} & = & \{\sigma^{ij},H\}\,,
\end{eqnarray}
where $\pi^{ij}$ and $\sigma^{ij}$ are the canonical momenta of the metric $\gamma_{ij}$ and $\phi_{ij}$ respectively, whereas $H$ is the total Hamiltonian of MTBG (given in Eq.\ (11) of \cite{DeFelice:2020ecp}). Then, for the considered ansatz, we can show that 
\begin{equation}
\sigma^{ij}=\alpha^2\,\pi^{ij}\,,
\end{equation}
together with
\begin{equation}
\{\sigma^{ij},H\}-\alpha^{2}\{\pi^{ij},H\}=[\dots]\,\gamma^{ij}=0\,,
\end{equation}
because the right hand side of this equation is proportional to Eq.\ (\ref{eq:defC}). Finally, the Hamilton equations of motion for the physical momenta $\pi^{ij}$ coincide with those of 
GR with cosmological constant.

In summary, what we find here is that any solutions of 
GR with cosmological constant
in a slicing satisfying the condition \eqref{VCDM_slicing},
such as a constant mean curvature slicing, are also solutions of the normal branch of MTBG (with untuned cosmological constants), provided that the two metric are parallel to each other (and $\lambda$, $\partial_i\bar\lambda$, 
and $\lambda^i$ all vanish) in the sense of \eqref{eq:norm_ansatz}.

As a consequence, if we consider a static and spherically symmetric background, the solutions which satisfy the imposed ansatz for MTBG, having $K=-3b_0=\textrm{constant}$, coincide with the general solutions given in \cite{DeFelice:2020onz} for a theory of minimally modified gravity, namely VCDM and VCCDM\footnote{Although there is no link between these theories (e.g.\ the number of the degrees of freedom are different), still it is interesting that they share same solutions on this background. In the following we also keep the same notation of \cite{DeFelice:2020onz} for an immediate comparison.} \cite{DeFelice:2020cpt,DeFelice:2020prd,DeFelice:2021xps,Maeda-san}, which can now be written as
\begin{eqnarray}
  g_{\mu\nu}\,dx^\mu\,dx^\nu&=&-\frac{N_0^2}{F(r)^2}\,dt^2+\left[F(r)\,dr+\left(b_0r-\frac{\kappa_0}{r^2}\right)N_0\,dt\right]^2+r^2
\left(
 d\theta^2
+\sin^2\theta d\varphi^2
\right),
\\
  f_{\mu\nu}\,dx^\mu\,dx^\nu&=&C_0^2\left\{-\frac{N_0^2}{F(r)^2}\,dt^2+\left[F(r)\,dr+\left(b_0r-\frac{\kappa_0}{r^2}\right)N_0\,dt\right]^2
+r^2\,
\left(
 d\theta^2
+\sin^2\theta d\varphi^2
\right)
\right\}\,,
  \end{eqnarray}
where $N_0$, $b_0$ and $\kappa_0$ are free constants\footnote{The constant parameter $N_0$, because of time-reparametrization invariance, can be safely set to unity.}, whereas
  \begin{equation}
    \frac1{F(r)^2}=1-\frac{2 \mu_0}{r}-\frac13\,\Lambda_0\,r^2+\frac{\kappa_0^2}{r^4}\,,\quad{\rm with}\quad \Lambda_0=\Lambda_{\rm eff}-3b_0^2\,.
  \end{equation}
  Several properties of the solutions were already discussed in
  \cite{DeFelice:2020onz}, e.g.\ they represent the Schwarzschild-de
  Sitter solution written in a constant-$K$ (in space and time)
  slicing, so that we here would like to discuss instead this solution in the
  context of MTBG.
We note that the case of $b_0=\kappa_0=0$
corresponds to a pair of the Schwarzschild-de Sitter spacetimes
sharing the same ADM mass and effective cosmological constant
written in the Schwarzschild coordinates.

  Then, although this class of solutions do exist in the normal branch and
  they can successfully provide a description for a BH, 
still we have to elaborate on the choice, in MTBG, for the chosen
  ansatz.
Indeed, both metrics $f_{\mu\nu}$ and $g_{\mu\nu}$ are parallel
  to each other.
This choice is legitimate, but we should ask ourselves
  whether it is physically motivated or not, more explicitly,
if this configuration is fine-tuned or not.
For instance, we know that this strong condition for the two metrics was not necessary in
  the self-acceleration branch. Furthermore, it is not clear whether a
  generic collapse in the normal branch would end up in this parallel
  configuration or not.

  Setting this problem for a moment aside, this solution for the normal branch was found by assuming the ansatz given in Eq.\ \eqref{eq:norm_ansatz},
  in particular a proportionality between the fiducial and the physical metrics on the
  background, namely $f_{\mu\nu}=C_0^2g_{\mu\nu}$, 
provided that a slicing 
satisfying $D_iD^iK=0$ exists, 
and the condition determining $C_0$ written in Eq.\ (\ref{eq:defC}) holds true. 
In fact,
  this same procedure can be successfully extended to other
  vacuum backgrounds provided a constant-$K$ slicing exists, such as the
  Kerr-de Sitter solution written in Boyer–Lindquist coordinates.

Once more, although an existence proof of non-perturbative solutions is now given in the normal branch of MTBG --- a result which is not trivial given the existence of constraints --- how to make sense of this parallel configuration for the two metrics remains. In this case as well, it would be interesting to either understand the collapse dynamics, or at least extend the solution by e.g.~perturbatively detuning the condition \eqref{VCDM_slicing}. Whether more generic solutions exist in the normal branch (in particular for the case $\lambda\neq0$), or the constraint equations are too restrictive and render the normal branch unviable for realistic situations remains to be studied.

\section{Static and spherically symmetric GR solutions with two distinct matter sectors}
\label{sec4}

We then introduce the matter sector $\Psi$,
which is divided into the two distinct sectors $\Psi_{m,g}$ and $\Psi_{m,f}$
minimally coupled to the physical and fiducial metrics $g_{\mu\nu}$ and $f_{\mu\nu}$
separately,
\begin{eqnarray}
\label{matt_action}
{\cal L}_m
\left[
N,N^i,\gamma_{ij};
M,M^i,\phi_{ij};
\Psi
\right]
=
{\cal L}_{m,g}
\left[
N,N^i,\gamma_{ij};\Psi_{m,g}
\right]
+
{\cal L}_{m,f}
\left[
M,M^i,\phi_{ij};
\Psi_{m,f}
\right].
\end{eqnarray}

\subsection{Static and spherically symmetric GR solutions with matter}

Before studying MTBG, we first consider GR with 
the gravitational constant $\kappa^2$
and 
the cosmological constant $\Lambda$
and assume that the metric $g_{\mu\nu}$ represents
a static and spherically symmetric solution in GR,
which is written in terms 
of the spatially-flat GP coordinates as
\begin{eqnarray}
\label{gp_coord}
g_{\mu\nu}dx^\mu dx^\nu
=
-A_0(r)dt^2
+\left(dr+N^r(r)dt\right)^2
+r^2\left(d\theta^2+\sin^2\theta d\varphi^2\right).
\end{eqnarray} 
We also assume that
$g_{\mu\nu}$ satisfies the Einstein equations in GR
with the matter energy-momentum tensor 
\begin{eqnarray}
\label{perfect1}
T_{(m,g)\mu\nu}
:= -\frac{2}{\sqrt{-g}}\frac{\delta {\cal L}_{m,g}}
            {\delta g^{\mu\nu}}
=
\left(\rho+p\right)u_\mu u_\nu
+p g_{\mu\nu},
\end{eqnarray}
where 
$u^{\mu}=\left(\frac{1}{\sqrt{A_0(r)-\left(N^r(r)\right)^2}},0,0,0\right)$
represents the four velocity,
and $\rho$ and $p$ are the energy density and pressure
of the perfect fluid, respectively. 
The Einstein equations are explicitly given by 
\begin{eqnarray}
\label{components_einstein1}
&&
-\frac{N^r(r)^2}{r^2A_0(r)}
+\frac{N^r(r)^2 A_0'(r)}{r A_0(r)^2}
-\frac{2N^r (r)N^r{}'(r) }{rA_0(r)}
=-\kappa^2 \rho(r)-\Lambda,
\nonumber\\
&&
-\frac{N^r(r)^2}{r^2A_0(r)}
+\frac{A_0'(r)}{rA_0(r)}
-\frac{2N^r (r)N^r{}'(r) }{rA_0(r)}
=\kappa^2 p(r)-\Lambda,
\nonumber\\
&&
  \frac{A_0'(r)}{2rA_0(r)}
+\frac{N^r(r)^2A_0'(r)}{2rA_0(r)^2}
-\frac{A_0'(r)^2}{4A_0(r)^2}
-\frac{2N^r(r)N^r{}'(r)}{rA_0(r)}
+\frac{N^r(r) A_0'(r) N^r{}'(r)}{2A_0(r)^2}
-\frac{N^r{}'(r)^2}{A_0(r)}
\nonumber\\
&& \quad + 
\frac{A_0''(r)}{2A_0(r)}
-\frac{N^r(r)N^r{}''(r)}{A_0(r)}
=\kappa^2 p(r)-\Lambda.
\end{eqnarray}

\subsection{Static and spherically symmetric GR solutions in MTBG}

In MTBG, we assume that the physical and fiducial metrics are given by Eq.\ \eqref{generalfg} 
with the spatially-flat form
\begin{eqnarray}
\label{solution_both_constant1}
&&
A_1(r)=1,
\quad 
A_{2}(r)=1,
\quad 
A_{1f}(r)=1,
\quad 
A_{2f}(r)=1.
\end{eqnarray}
We note that the constraint \eqref{const} is trivially satisfied
for the spatially-flat coordinates \eqref{solution_both_constant1}.
We also assume that the matter in the physical and fiducial sectors \eqref{matt_action}
is given by the perfect fluids with the energy-momentum tensors,
Eq.\ \eqref{perfect1}
and
\begin{eqnarray}
\label{perfect2}
T_{\mu\nu}^{(m,f)}
:= -\frac{2}{\sqrt{-f}}\frac{\delta {\cal L}_{m,f}}{\delta f^{\mu\nu}}
=
\left(\rho_f+p_f\right)u_{(f)\mu} u_{(f)\nu}
+p_f f_{\mu\nu},
\end{eqnarray}
with the matter four velocity
$u_{(f)\mu}=\left(\frac{1}{C_0 \sqrt{A_{0f}(r)-\left(N_f^r(r)\right)^2}},0,0,0\right)$,
$\rho_f$, and $p_f$
being the four velocity,
energy density,
and pressure
of the perfect fluid, respectively, 
in the fiducial sector.
The coefficients in
the constraint equations \eqref{cons2} and \eqref{cons3}
are given by 
\begin{eqnarray}
\alpha_{1\gamma}
&=&
\frac{(C_0^2 c_1+2C_0 c_2+c_3) m^2}
       {r\sqrt{A_0(r)}}
\left[
r N^r{}'(r)+2N^r(r)
\right],
\quad 
\alpha_{1\phi}
=
-
\frac{\left(C_0^2c_1+2C_0 c_2+c_3\right) m^2}
       {C_0^3r\sqrt{A_{0f} (r)} }
\left[
r N_f^r{}'(r)
+2N_f^r(r)
\right],
\nonumber\\
\alpha_{2\gamma}
&=&
\frac{3\left(C_0^2 c_1+2C_0 c_2+c_3\right)^2m^4}
       {8\sqrt{A_0(r)}},
\quad 
\alpha_{2\phi}
=
\frac{3\left(C_0^2 c_1 +2C_0 c_2+c_3\right)^2m^4}
       {8\tilde{\alpha}^2C_0^5\sqrt{A_{0f}(r)}}.
\end{eqnarray}
We choose the self-accelerating branch \eqref{additional}~\footnote{We choose the self-accelerating branch since, as in the vacuum case discussed in Sec.~\ref{sec3}, there is no consistent solution for static and spherically symmetric stars compatible with the constraint conditions satisfying the equivalent of Eq.~\eqref{fgzero}.} 
and impose the conditions \eqref{cond}.
We also assume that 
the physical and fiducial metrics satisfy the Einstein equations
in GR coupled to
the matter energy-momentum tensors \eqref{perfect1} and \eqref{perfect2}
respectively,
which are explicitly given by Eq.\ \eqref{components_einstein1}
in the physical frame
and 
\begin{eqnarray}
\label{components_einstein2}
&&
-\frac{N_f^r(r)^2}{r^2A_{0f}(r)}
+\frac{N_f^r(r)^2 A_{0f}'(r)}{r A_{0f}(r)^2}
-\frac{2N_f^r (r)N_f^r{}'(r) }{rA_{0f}(r)}
=
C_0^2
\left(-\frac{\kappa^2}{{\tilde \alpha}^2} \rho_f(r)-\Lambda_f\right),
\nonumber\\
&&
-\frac{N_f^r(r)^2}{r^2A_{0f}(r)}
+\frac{A_{0f}'(r)}{rA_{0f}(r)}
-\frac{2N_f^r (r)N_f^r{}'(r) }{rA_{0f}(r)}
=
C_0^2
\left(
\frac{\kappa^2}{{\tilde\alpha}^2} p_f(r)-\Lambda_f
\right),
\nonumber\\
&&
  \frac{A_{0f}'(r)}{2rA_{0f}(r)}
+\frac{N_f^r(r)^2A_{0f}'(r)}{2rA_{0f}(r)^2}
-\frac{A_{0f}'(r)^2}{4A_{0f}(r)^2}
-\frac{2N_f^r(r)N_f^r{}'(r)}{rA_{0f}(r)}
+\frac{N_f^r(r) A_{0f}'(r) N_f^r{}'(r)}{2A_{0f}(r)^2}
-\frac{N^r{}'(r)^2}{A_{0f}(r)}
\nonumber\\
&& \quad + 
 \frac{A_{0f}''(r)}{2A_{0f}(r)}
-\frac{N_f^r(r)N_f^r{}''(r)}{A_{0f}(r)}
=
C_0^2
\left(
\frac{\kappa^2}{{\tilde \alpha}^2} p_f(r)-\Lambda_f
\right),
\end{eqnarray}
in the fiducial frame.
Substituting all the metric conditions into the metric and constraint equations
in MTBG,
we find the solution~\eqref{all_lambda_vanish}.
Thus,
as well as the vacuum case,
the self-accelerating branch in MTBG
with Eqs.\ \eqref{cond} and \eqref{additional}
allows the static and spherically symmetric GR solutions
with matter fields.

\subsection{Regularity at the center of the star}

We check the regularity at the center of the star.
Since the four-dimensional diffeomorphism invariance of the spacetime
is explicitly broken in MTBG in the unitary gauge,
even if two metrics in a coordinate system
describe a solution in MTBG,
these expressed in another coordinate system
may not be a solution.
Thus, if a coordinate singularity exists in one coordinate system where MTBG allows a solution,
it may be a physical singularity in contrast to the case of GR where a coordinate singularity is removed via a coordinate transformation.

A regular static and spherically symmetric metric
written in the Schwarzschild coordinates,
$g_{\mu\nu} dx^\mu dx^\nu=-f(T,r)dT^2+\frac{dr^2}  {1-\frac{2m(T,r)}{r}}+r^2d\Omega^2$,
can be brought to
the spatially-flat form \cite{DeFelice:2018vza}
\begin{eqnarray}
g_{\mu\nu}
dx^\mu 
dx^\nu
=
-
\frac{f}{1-\frac{2m}{r}}
dt^2
+
\left(
dr
-\frac{\sqrt{2mf}}
         {\sqrt{r-2m}}
dt
\right)^2
+
r^2
\left(
d\theta^2+\sin^2\theta d\varphi^2
\right),
\end{eqnarray}
by the coordinate transformation $dT=dt+\sqrt{\frac{2m}{f(r-2m)}} dr$.
For the regularity
in the spatially-flat coordinates,
we require the regularity and isotropy
at the center of the star $r=0$ in both sectors.
Computing the traceless part of the extrinsic curvature tensors
\eqref{extrinsic},
we obtain 
\begin{eqnarray}
K^r{}_r-\frac{1}{3}K
=\frac{2}{3N}
\left(N^r_{,r}-\frac{N^r}{r}
\right),
\quad 
N^r
=
-\frac{\sqrt{2mf}}
         {\sqrt{r-2m}}.
\end{eqnarray}
Assuming a stellar solution in each sector
whose metric and matter field are regular 
at the center of the star $r=0$ in the Schwarzschild coordinates,
the leading behavior around $r=0$ are given by 
$f(r)=f_0 +f_2 r^2+{\cal O} (r^4)$,
and 
$m(r)=  m_3r^3+{\cal O} (r^5)$
where 
$f_0$, $f_2$, and $m_3$ are constants
and hence
$K^r{}_r-\frac{1}{3}K\sim r^2$
in the spatially-flat coordinates.
Thus, 
the extrinsic curvature is regular and isotropic in the physical frame.
Since the two sectors are decoupled,
the same conclusion also holds in the fiducial sector.

\section{Solutions in the small $m^2$ expansion}
\label{sec5}

In this section, 
we construct static and spherically symmetric vacuum solutions
perturbatively in the small mass limit of MTBG, $m\to 0$.
We regard the graviton mass squared $m^2$ as an expansion parameter,
and expand the metric variables
and 
the Lagrange multipliers.
We expect that
at the leading order ${\cal O} (m^0)$ 
the Schwarzschild-de Sitter metric
written in the spatially-flat GP coordinates
with the cosmological constants
$V$ and $V_f$
and the masses $M$ and $M_f$
in the physical and fiducial frames, 
respectively,
are realized,
and
at the next-to-leading order ${\cal O} (m^2)$
the nontrivial corrections to the GR solutions
may be obtained.
We investigate
whether and how
the finite nonzero values $m^2$
can provide the deviation from 
the Schwarzschild-de Sitter metrics in GR in each sector.

We modify the precursor action \eqref{lag2},
so that 
the GR part of the action also contains the cosmological constants
$V$ and $V_f$ 
in the physical and fiducial sectors,
respectively, as
\begin{eqnarray}
\label{pre_2}
{\cal L}_{\rm pre}
&:=&
\sqrt{-g}
\left(R[g]-2V\right)
+
\tilde{\alpha}^2
\sqrt{-f} 
\left(R[f]
-2V_f
\right)
-
m^2
\left(
N\sqrt{\gamma}{\cal H}_0
+
M\sqrt{\phi}\tilde{\cal H}_0
\right),
\end{eqnarray}
where $V$ and $V_f$ are assumed to be positive constants.
We extend the metric ansatz \eqref{generalfg}
to the case of the spherically symmetric but time-dependent spacetimes,
where all functions $A_0$, $A_{0f}$, $N^r$, $N_f^r$,
$A_1$, $A_{1f}$, $A_2$, and $A_{2f}$
in Eq.~\eqref{generalfg}
are promoted to the functions of $t$ as well as $r$,
such as $A_0(r)\to A_0 (t,r)$.
Substituting the time-dependent and spherically symmetric
metrics of the physical and fiducial sectors
into the action \eqref{action}
and
varying it with respect to 
$A_0$, $A_{0f}$, $N^r$, $N_f^r$,
$A_1$, $A_{1f}$, $A_2$, and $A_{2f}$,
respectively,
we obtain 
the Euler-Lagrange equations 
for these metric components.
We also take the constraint equations
\eqref{cons2}, \eqref{cons3}, and \eqref{const}
into consideration.
Because of the spherical symmetry of the background spacetimes,
we may set $\lambda^\theta=\lambda^\varphi=0$
as Eq.\ \eqref{lambda_a_vanish}.
We also promote the nontrivial Lagrange multipliers
$\lambda$, ${\bar \lambda}$, and ${\lambda}^r$
to functions of $(t,r)$.

We then expand the metric variables 
with respect to
the Schwarzschild-de Sitter metrics
written in the spatially-flat coordinates
in terms of $m^2$ as
\begin{eqnarray}
\label{perturbative_expansion}
A_0(t,r)&=&1+m^2 a_0(t,r)+{\cal  O}\left(m^4\right),
\quad 
N^r(t,r)
=
\pm
\sqrt{
\frac{2M}{r}
+\frac{V}{3}r^2}
+m^2 n^r(t,r)
+{\cal O} (m^4),
\nonumber\\ 
A_1(t,r)&=&1+m^2 a_1(t,r)+ {\cal O}\left(m^4\right),
\quad
A_2(t,r)=1+m^2 a_2(t,r)+{\cal  O}\left(m^4\right),
\nonumber\\
A_{0f}(t,r)&=&1+m^2 a_{0f}(t,r)+{\cal  O}\left(m^4\right),
\quad 
N_f^r(t,r)
=
\pm
\sqrt{
\frac{2M_f}{r}
+\frac{C_0^2V_f}
         {3}
r^2}
+m^2 n_f^r(t,r)
+{\cal O} (m^4),
\nonumber\\
A_{1f}( t,r)&=&1+m^2 a_{1f}(t,r)+ {\cal O}\left(m^4\right),
\quad 
A_{2f}(t,r)=1+m^2 a_{2f}(t,r)+{\cal  O}\left(m^4\right),
\end{eqnarray}
where
$a_0(t,r)$, $n^r(t,r)$, $a_1(t,r)$, $a_2(t,r)$,
$a_{0f}(t,r)$, $n^r_f(t,r)$, $a_{1f}(t,r)$, and $a_{2f}(t, r)$
are functions of $t$ and $r$. 
We also expand the Lagrange multipliers 
$\lambda(t,r)$, ${\bar \lambda} (t,r)$, and $\lambda^r(t,r)$ 
in terms of $m^2$ as  
\begin{eqnarray}
\label{lag_ansatz2}
\lambda(t,r)&=&
\frac{\Lambda_{-2}(t,r)}{m^2}
+\Lambda_0(t,r)
+{\cal O} (m^2),
\nonumber\\
{\bar \lambda}(t,r)&=&
\frac{ {\bar \Lambda}_{-2}(t,r)}{m^2}
+{\bar \Lambda}_0(t,r)
+{\cal O} (m^0),
\nonumber\\
\lambda^r(t,r)
&=&
\frac{
 \Lambda^r_{-2}(t,r)}{m^2}
+
 \Lambda^r_{0}(t,r)
+{\cal O} (m^2),
\end{eqnarray}
where the leading dependence on $m$ is chosen as $m^{-2}$,
since the Lagrange multipliers always appear with $m^2$ in the Lagrangian \eqref{lag2}
and may in principle contribute nontrivially at ${\cal O} (m^0)$.
We consider the radial component of the constraint \eqref{const}.
At ${\cal O} (m^2)$,
the radial component of Eq.~\eqref{const}
is satisfied for 
\begin{eqnarray}
\left(
(c_2+C_0 c_1)+\beta (c_3+c_2C_0)
\right)
\left[
  a_1(t,r)
-a_2(t,r)
-a_{1f}(t,r)
+a_{2f}(t,r)
-r a_{2,r}(t,r)
+r a_{2f,r} (t,r)
\right]
=0,
\end{eqnarray}
which provides the following three cases

\begin{enumerate}

\item\label{case1}
$a_{1f}(t,r)
=
  a_1(t,r)
-a_2(t,r)
+a_{2f}(t,r)
-r a_{2,r}(t,r)
+r a_{2f,r} (t,r)
$,

\item
$(c_2+C_0 c_1)+\beta (c_3+c_2C_0)=0$ with $\beta C_0 \neq 1$,
\label{case2}

\item
$(c_2+C_0 c_1)+\beta (c_3+c_2C_0)=0$ with $\beta C_0= 1$,
 yielding the degenerate condition \eqref{additional}.
\label{case3}

\end{enumerate}
In order to satisfy the constraint equations \eqref{cons2} and \eqref{cons3},
we consider the self-accelerating branch (See Appendix \ref{app:nogo-normalbranch} for the corresponding analysis of the normal branch). The self-accelerating branch is given by Eq.\ \eqref{additional} which automatically satisfies the constraint equations \eqref{cons2} and \eqref{cons3}.\

In Case \ref{case1},
at  ${\cal O} (m^2)$,
the Euler-Lagrange equations
provide the solution for the metrics
\begin{eqnarray}
\label{sol_sa_1}
a_0(t,r)
&=&
q_{n_0}(t)
-b_1 (r),
\nonumber\\
a_1(t,r)
&=&
b_1(r)
+a_2(t,r) 
+ r a_{2,r}(t,r),
\nonumber
\\
a_{0f}(t,r)
&=&
q_{n_{0f}} (t)
-b_{1}(r),
\nonumber\\
a_{1f}(t,r)
&=&
b_{1}(r)
+a_{2f}(t,r) 
+ r a_{2f,r}(t,r),
\nonumber\\
n^r(t,r)
&=&
\pm
\Big[
\frac{\sqrt{3}}       
        {12\sqrt{r\left(6M +r^3V\right)}}
\nonumber\\
&\times&
\left(
r^3\left(-2C_0^3c_1-3C_0^2 c_2+c_4\right)
-2 \left(6M+3r+r^3V\right)b_1(r)
-18M a_2 (t,r)
-2r\left(6M+r^3V \right) a_{2,r}(t,r)
\right)
\nonumber\\
&+ &
\frac{\sqrt{3} M q_{n^r}}
        {\sqrt{r\left(6M+r^3V\right)}}
+ 
\frac{q_{n_0}(t) }{2}
\sqrt{
\frac{2M} {r}
+\frac{V}{3}r^2}
\Big]
+\frac{r}{2}a_{2,t} (t,r),
\nonumber\\
n_f^r(t,r)
&=&
\pm 
\Big[
\frac{\sqrt{3}}
        {12 {\tilde\alpha}^2 \sqrt{r\left(6M_f +r^3C_0^2V_f\right)}}
\nonumber\\
&\times&
\left(
r^3\left(C_0^2c_0+2C_0 c_1+c_2\right)
-2{\tilde\alpha}^2 (6M_f+3r+C_0^2r^3 V_f) b_{1}(r)
-18\tilde{\alpha}^2M_f a_{2f} (t,r)
-2{\tilde \alpha}^2r\left(6M_f+C_0^2 r^3V_f \right) a_{2f,r}(t,r)
\right)
\nonumber\\
&+&
\frac{\sqrt{3} M_f q_{n_f^r}}
        {\sqrt{r\left(6M_f +r^3 C_0^2V_f\right)}}
+
\frac{q_{n_{0f}} (t) }{2}
\sqrt{
\frac{2M_f}{r}
+\frac{C_0^2V_f}
         {3}
r^2}
\Big]
+\frac{r}{2}a_{2f,t} (t,r),
\end{eqnarray}
and 
that 
for the Lagrange multipliers
\begin{eqnarray}
\label{setting_lambda}
&&
\Lambda_{-2}(t,r)
=0,
\quad 
{\bar\Lambda}_{-2} (t,r)
=
d_0(t),
\quad 
\Lambda^r_{-2}(t,r)
=0,
\nonumber
\\
&&
\Lambda_{0}(t,r)
=0,
\quad 
{\bar\Lambda}_{0} (t, r)
=
e_0(t)
+
\frac{e_1(t)}{r},
\quad 
\Lambda^r_{0}(t,r)
=0,
\end{eqnarray}
where
$\left(q_{n_0}(t),q_{n_{0f}} (t),d_0(t), e_0(t), e_1 (t)\right)$
are pure functions of $t$,
$b_1(r)$ is a pure function of $r$,
and 
$\left(q_{n^r},q_{n_f^r}\right)$
are integration constants, 
respectively.
We note
that
the coefficients 
$e_0(t)$ and $e_1(t)$ 
are those for the solutions of the Laplace equation in the Euclid space
and do not play a physical role.
Up to ${\cal O} (m^2)$,
$a_2(t,r)$ and $a_{2f}(t,r)$
are not determined. 
The solution describes the Schwarzschild-de Sitter metric
with the effective cosmological constants and masses
\begin{eqnarray}
\label{sch_ds_parameters}
V_{\rm eff}
&:=&
V
+
\frac{m^2}{2}
\left(
-2C_0^3 c_1
-3 C_0^2 c_2
+c_4
\right),
\qquad 
V_{f, {\rm eff}}
:=
V_f
+
\frac{m^2}{2{\tilde\alpha}^2}
\left(
  c_0
+2c_1 C_0^{-1}
+c_2 C_0^{-2}
\right),
\nonumber
\\
M_{\rm eff}
&:=&
M
\left(
1+m^2 q_{n^r}
\right),
\qquad 
M_{f,{\rm eff}}
:=
M_f
\left(
1+m^2 q_{n^r_f}
\right),
\end{eqnarray}
respectively,
written in the non-standard coordinates.
$q_{n^r}$ and $q_{n^r_f}$ correspond to the constant shifts of masses,
which can be absorbed into a redefinition of $M$ and $M_f$.
The functions $q_{n_0}(t)$ and $q_{n_{0f}}(t)$
correspond to the reparametrization of time
in the physical and fiducial sectors,
$dt\to \left(1+\frac{m^2}{2} q_{n_0}(t)\right)dt$
and  
$dt\to \left(1+\frac{m^2}{2} q_{n_{0f}}(t)\right)dt$,
respectively,
and 
only 
one of them can be set to vanish.
Similarly, 
up to ${\cal O} (m^2)$, 
the functions $a_2(t,r)$ and $a_{2f} (t,r)$ correspond to 
the reparametrization of the radial coordinates 
$r\to r \left(1+\frac{m^2}{2}a_2(t,r)\right)$
and 
$r\to r \left(1+\frac{m^2}{2}a_{2f}(t,r)\right)$.

In Case \ref{case2} with $\beta C_0\neq 1$,
the compatibility with the self-accelerating condition \eqref{additional} yields
\begin{eqnarray}
&&
c_2=-C_0 c_1,
\quad
c_3=C_0^2 c_1.
\end{eqnarray}
At ${\cal O} (m^2)$,
the Euler-Lagrange equations 
provide 
the solution for the metrics
\begin{eqnarray}
a_0(t,r)
&=&
q_{n_0}(t)
-b_1 (r),
\nonumber
\\
a_1(t,r)
&=&
b_1(r)
+a_2(t,r) 
+ r a_{2,r}(t,r),
\nonumber
\\
a_{0f}(t,r)
&=&
q_{n_{0f}} (t)
-b_{1 f}(r),
\nonumber\\
a_{1f}(t,r)
&=&
b_{1f}(r)
+a_{2f}(t,r) 
+ r a_{2f,r}(t,r),
\nonumber\\
n^r(t,r)
&=&
\pm
\Big[
\frac{\sqrt{3}}       
        {12\sqrt{r\left(6M +r^3V\right)}}
\nonumber\\
&\times&
\left(
r^3\left(C_0^3c_1+c_4\right)
-2 \left(6M+3r+r^3V\right)b_1(r)
-18M a_2 (t,r)
-2r\left(6M+r^3V \right) a_{2,r}(t,r)
\right)
\nonumber\\
&+ &
\frac{\sqrt{3} M q_{n^r}}
        {\sqrt{r\left(6M+r^3V\right)}}
+ 
\frac{q_{n_0}(t) }{2}
\sqrt{
\frac{2M} {r}
+\frac{V}{3}r^2}
\Big]
+\frac{r}{2}a_{2,t} (t,r),
\nonumber\\
n_f^r(t,r)
&=&
\pm 
\Big[
\frac{\sqrt{3}}
        {12 {\tilde\alpha}^2 \sqrt{r\left(6M_f +r^3C_0^2V_f\right)}}
\nonumber\\
&\times&
\Big(
r^3
\left(C_0^2c_0+C_0 c_1\right)
-2{\tilde\alpha}^2 (6M_f+3r+C_0^2r^3 V_f) b_{1f}(r)
-18\tilde{\alpha}^2M_f a_{2f} (t,r)
-2{\tilde \alpha}^2r\left(6M_f+C_0^2 r^3V_f \right) a_{2f,r}(t,r)
\Big)
\nonumber\\
&+&
\frac{\sqrt{3} M_f q_{n_f^r}}
        {\sqrt{r\left(6M_f +r^3 C_0^2V_f\right)}}
+
\frac{q_{n_{0f}} (t) }{2}
\sqrt{
\frac{2M_f}{r}
+\frac{C_0^2V_f}
         {3}
r^2}
\Big]
+\frac{r}{2}a_{2f,t} (t,r),
\end{eqnarray}
and that for the Lagrange multipliers
\begin{eqnarray}
&&
\Lambda_{-2}(t,r)
=0,
\quad 
{\bar\Lambda}_{-2} (t,r)
=
e_0(t)
+
\frac{e_1(t)}{r},
\quad 
\Lambda^r_{-2}(t,r)
=0,
\end{eqnarray}
where
$\left(q_{n_0}(t),q_{n_{0f}} (t),d_0(t), e_0(t), e_1 (t)\right)$
are pure functions of $t$,
$\left(b_1(r),b_{1f}(r)\right)$
are pure functions of $r$,
and 
$\left(q_{n^r},q_{n_f^r}\right)$
are integration constants, 
respectively.
We note
that
the coefficients 
$e_0(t)$ and $e_1(t)$ 
are those for the solutions of the Laplace equation in the Euclid space
and do not play a physical role.
The functions
$a_2(t,r)$,
$a_{2f}(t,r)$,
$\Lambda_0(t,r)$,
$ {\bar \Lambda}_0(t,r)$,
and
${\Lambda}^r_0(t,r)$
are not determined
by the Euler-Lagrange equations 
up to ${\cal O} (m^2)$.
At the next order of ${\cal O} (m^6)$,
the constraint \eqref{const} yields 
\begin{eqnarray}
b_1(r)=b_{1f}(r),
\quad
\text{or}
\quad 
a_{2}(t,r)=a_{2f}(t,r).
\end{eqnarray}
The solution describes the Schwarzschild-de Sitter metrics
with the effective cosmological constants and masses
\begin{eqnarray}
V_{\rm eff}
&:=&
V
+
\frac{m^2}{2}
\left(C_0^3c_1+c_4\right),
\qquad 
V_{f, {\rm eff}}
:=
V_f
+
\frac{m^2}{2{\tilde\alpha}^2}
\left(C_0^2c_0+C_0 c_1\right),
\nonumber
\\
M_{\rm eff}
&:=&
M
\left(
1+m^2 q_{n^r}
\right),
\qquad 
M_{f,{\rm eff}}
:=
M_f
\left(
1+m^2 q_{n^r_f}
\right),
\end{eqnarray}
respectively,
written in the non-standard coordinates.
On the other hand, the functions $b_1(r)$ and $b_{1f}(r)$ correspond to the freedom to choose radial coordinates for $g_{\mu\nu}$ and $f_{\mu\nu}$, respectively. 

In Case \ref{case3},
all the constraints \eqref{cons2}, \eqref{cons3}, and \eqref{const}
are satisfied under the same condition
as Eq.~\eqref{additional}.
At ${\cal O} (m^2)$,
the Euler-Lagrange equations
provide
the solution for the metrics
\begin{eqnarray}
\label{sol_sa_3}
a_0(t,r)
&=&
q_{n_0}(t)
-b_{1} (r),
\nonumber\\
a_1(t,r)
&=&
b_1(r)
+a_2(t,r) 
+ r a_{2,r}(t,r),
\nonumber
\\
a_{0f}(t,r)
&=&
q_{n_{0f}} (t)
-b_{1f}(r),
\nonumber\\
a_{1f}(t,r)
&=&
b_{1f}(r)
+a_{2f}(t,r) 
+ r a_{2f,r}(t,r),
\nonumber\\
n^r(t,r)
&=&
\pm
\Big[
\frac{\sqrt{3}}       
        {12\sqrt{r\left(6M +r^3V\right)}}
\nonumber\\
&\times&
\left(
r^3\left(-2C_0^3c_1-3C_0^2 c_2+c_4\right)
-2 \left(6M+3r+r^3V\right)b_1(r)
-18M a_2 (t,r)
-2r\left(6M+r^3V \right) a_{2,r}(t,r)
\right)
\nonumber\\
&+ &
\frac{\sqrt{3} M q_{n^r}}
        {\sqrt{r\left(6M+r^3V\right)}}
+ 
\frac{q_{n_0}(t) }{2}
\sqrt{
\frac{2M} {r}
+\frac{V}{3}r^2}
\Big]
+\frac{r}{2}a_{2,t} (t,r),
\nonumber\\
n_f^r(t,r)
&=&
\pm 
\Big[
\frac{\sqrt{3}}
        {12 {\tilde\alpha}^2 \sqrt{r\left(6M_f +r^3C_0^2V_f\right)}}
\nonumber\\
&\times&
\left(
r^3\left(C_0^2c_0+2C_0 c_1+c_2\right)
-2{\tilde\alpha}^2 (6M_f+3r+C_0^2r^3 V_f) b_{1f}(r)
-18\tilde{\alpha}^2M_f a_{2f} (t,r)
-2{\tilde \alpha}^2r\left(6M_f+C_0^2 r^3V_f \right) a_{2f,r}(t,r)
\right)
\nonumber\\
&+&
\frac{\sqrt{3} M_f q_{n_f^r}}
        {\sqrt{r\left(6M_f +r^3 C_0^2V_f\right)}}
+
\frac{q_{n_{0f}} (t) }{2}
\sqrt{
\frac{2M_f}{r}
+\frac{C_0^2V_f}
         {3}
r^2}
\Big]
+\frac{r}{2}a_{2f,t} (t,r),
\end{eqnarray}
and that for the Lagrange multipliers
\begin{eqnarray}
&&
\Lambda_{-2}(t,r)=0,
\qquad 
{\bar \Lambda}_{-2}(t,r)
=d_0(t),
\qquad 
{\Lambda}^r_{-2}(t,r)
=0,
\nonumber\\
&&
\Lambda_{0}(t,r)=0,
\qquad
{\bar \Lambda}_{0}(t,r)
=
e_0(t)
+
\frac{e_1(t)}{r},
\end{eqnarray}
respectively,
where
$\left(q_{n_0}(t),q_{n_{0f}} (t),d_0(t), e_0(t), e_1 (t)\right)$
are pure functions of $t$,
$\left(b_1(r),b_{1f}(r)\right)$ are pure functions of $r$,
and 
$\left(q_{n^r},q_{n_f^r}\right)$
are integration constants, 
respectively.
We note
that
the coefficients 
$e_0(t)$ and $e_1(t)$ 
are those for the solutions of the Laplace equation in the Euclid space
and do not play a physical role.
We also note that
$a_2(t,r)$,
$a_{2f}(t,r)$,
${\Lambda}^r_{-2}(t,r)$,
and
${\Lambda}^r_{0}(t,r)$
are not determined
by the Euler-Lagrange equations 
up to ${\cal O} (m^2)$.
The solution describes the Schwarzschild-de Sitter metrics
with the effective cosmological constants and masses
\begin{eqnarray}
V_{\rm eff}
&:=&
V
+
\frac{m^2}{2}
\left(
-2C_0^3 c_1
-3 C_0^2 c_2
+c_4
\right),
\qquad 
V_{f, {\rm eff}}
:=
V_f
+
\frac{m^2}{2{\tilde\alpha}^2}
\left(
  c_0
+2c_1 C_0^{-1}
+c_2 C_0^{-2}
\right),
\nonumber
\\
M_{\rm eff}
&:=&
M
\left(
1+m^2 q_{n^r}
\right),
\qquad 
M_{f,{\rm eff}}
:=
M_f
\left(
1+m^2 q_{n^r_f}
\right),
\end{eqnarray}
respectively,
written in the non-standard coordinates.
On the other hand, the functions $b_1(r)$ and $b_{1f}(r)$ correspond to the freedom to choose radial coordinates for $g_{\mu\nu}$ and $f_{\mu\nu}$, respectively.

\section{Nonperturbative static and spherically symmetric vacuum solutions}

Having constructed the vacuum static and spherically symmetric solutions 
perturbatively in the small $m^2$ limit
in the self-accelerating branch \eqref{additional}~\footnote{In the normal branch the small $m^2$ expansion does not allow for the spatially-flat Schwarzschild-de Sitter metrics as the zero-th order solutions. See Appendix \ref{app:nogo-normalbranch} for details.},
the perturbative metrics agreed with the Schwarzschild-de Sitter metric
written in nonstandard coordinates.
This makes us expect
that beyond the perturbative construction
the Schwarzschild-de Sitter metric written 
in the nonstandard coordinates
should be a solution in
the self-accelerating branch of MTBG
for a generic value of $m^2$.
Here, we confirm this.

We start with the Schwarzschild-de-Sitter metrics
written in the Schwarzschild coordinates,
\begin{eqnarray}
\label{sch_ds_static}
g_{\mu\nu}dx^\mu dx^\nu
&=&
-
\left(
1
-\frac{2M}{r}
+\frac{V_{\rm eff}}{3}r^2
\right)
dT(t,r)^2
+
\frac{dr^2}
{1-\frac{2M}{r}
+\frac{V_{\rm eff}}{3}r^2}
+
r^2
\left(
d\theta^2
+\sin^2\theta d\varphi^2
\right),
\nonumber\\
f_{\mu\nu}dx^\mu dx^\nu
&=&
C_0^2
\left[
-
\left(
1
-\frac{2M_f}{r}
+\frac{C_0^2 V_{f,{\rm eff}}}{3}r^2
\right)
dT_f(t,r)^2
+
\frac{dr^2}
{1-\frac{2M_f}{r}
+\frac{C_0^2V_{f,{\rm eff}}}{3}r^2}
+
r^2
\left(
d\theta^2
+\sin^2\theta d\varphi^2
\right)
\right],
\end{eqnarray}
where the Schwarzschild times $T(t,r)$ and $T_f(t,r)$
in the physical and fiducial sectors, respectively, 
are related to the MTBG coordinates $(t,r)$ by 
\begin{eqnarray}
T(t,r):=\int h(t) dt +\int  g(r)dr,
\qquad 
T_f(t,r):=\int h_f(t) dt+ \int g_f(r)dr.
\end{eqnarray}
Here, 
$g(r)$ and $g_f(r)$ are pure functions of $r$,
and
$h(t)$ and $h_f(t)$ are pure functions of $t$.
The Schwarzschild-de Sitter metrics \eqref{sch_ds_static} are then written 
in the form of Eq.~\eqref{generalfg} by 
\begin{eqnarray}
\label{mtbg_nonstandard}
A_0(t,r)
&=&
\frac{3r\left(3r-6M-r^3V_{\rm eff}\right) h(t)^2}
        {9r^2-\left(3r-6M-r^3V_{\rm eff}\right)^2 g(r)^2},
\quad 
N^r(t,r)
=
-
\frac{\left(3r-6M-r^3V_{\rm eff}\right)^2 g(r) h(t)}
        {9r^2-\left(3r-6M-r^3V_{\rm eff}\right)^2 g(r)^2},
\nonumber\\
A_1(t,r)
&=&
\frac{h(t)^2}{A_0(t,r)},
\quad 
A_2(t,r)
=
1,
\nonumber
\\
A_{0f}(t,r)
&=&
\frac{3r\left(3r-6M_f-r^3C_0^2 V_{f,{\rm eff}}\right) h_f(t)^2}
        {9r^2-\left(3r-6M_f-r^3 C_0^2V_{f, {\rm eff}}\right)^2 g_f(r)^2},
\quad 
N_f^r(t,r)
=
-
\frac{\left(3r-6M_f-r^3 C_0^2 V_{f, {\rm eff}} \right)^2g_f(r) h_f(t)}
        {9r^2-\left(3r-6M_f-r^3C_0^2V_{f, {\rm eff}}\right)^2 g_f(r)^2},
\nonumber\\
A_{1f}(t,r)
&=&
\frac{h_f(t)^2}{A_{0f}(t,r)},
\quad 
A_{2f}(t,r)
=
1,
\end{eqnarray}
which describe the Schwarzschild-de Sitter solutions written in the nonstandard coordinates.
Choosing the self-accelerating branch \eqref{additional},
for $(c_2+C_0 c_1)+\beta (c_3+c_2C_0)\neq 0$
the radial component of the constraint equation \eqref{const} yields
\begin{eqnarray}
\label{gf}
g_f(r)
=
\mp
\frac{
\sqrt{
9r^2
\left(
-6M+6M_f+r^3\left(C_0^2 V_{f,{\rm eff}}-V_{\rm eff}\right)
\right)
+
\left(
3r-6M_f-r^3C_0^2V_{f, {\rm eff}}
\right)
\left(
3r-6M-r^3V_{\rm eff}
\right)^2
g(r)^2
}
}
        {\left(3r-6M_f-C_0^2r^3V_{f,{\rm eff}}\right)\sqrt{3r-6M-r^3V_{\rm eff}}},
\end{eqnarray}
which relates $g_f(r)$ and $g(r)$.
We note that $h(t)$ and $h_f(t)$ are not constrained at all
by the constraint and metric equations.
All the nontrivial components of the metric equations and the rest of the constraint equations are consistently satisfied 
for 
\begin{eqnarray}
\label{veff}
V_{\rm eff}
&:=&
V
+
\frac{m^2}{2}
\left(
-2C_0^3 c_1
-3 C_0^2 c_2
+c_4
\right),
\qquad 
V_{f, {\rm eff}}
:=
V_f
+
\frac{m^2}{2{\tilde\alpha}^2}
\left(
  c_0
+2c_1 C_0^{-1}
+c_2 C_0^{-2}
\right),
\label{veff_f}
\end{eqnarray}
together with the condition \eqref{additional},
which agree with 
Eq.\ \eqref{sch_ds_parameters}
as well as 
\begin{eqnarray}
\label{lambda_conditions_three}
\Lambda=0,
\qquad 
{\bar \Lambda}=d_0(t),
\qquad 
{\lambda}^r=0,
\end{eqnarray}
where $d_0(t)$ is a pure function of $t$.

We now show that
the perturbative solution 
can be recovered in the limit of small $m^2$.
We assume the perturbative expansion of the free function $g(r)$
in the $m\to 0 $ limit
\begin{eqnarray}
g(r)
&=&
g_0(r)
+
g_2(r) m^2
+
{\cal O} (m^4),
\nonumber\\
h(t)
&=&
1
+
\frac{m^2}{2}
q_{n_0}(t)
+
{\cal O} (m^4),
\qquad 
h_{f}(t)
=
1
+
\frac{m^2}{2}
q_{n_{0f}}(t)
+{\cal O} (m^4),
\end{eqnarray}
where $g_0(r)$ and $g_2(r)$ are pure functions of $r$,
and 
$q_{n_0}(t)$ and $q_{n_{0f}}(t)$
are functions of $t$, respectively.
Requiring that the ${\cal O} (m^0)$ part 
of the metrics
can be expressed by
the Schwarzschild solution in the 
spatially-flat coordinates,
we find
\begin{eqnarray}
g_0(r)
&=&
\mp
\frac{\sqrt{3(6Mr+r^4V)}}
                 {3r-6M-r^3V},
\nonumber\\
g_2(r)
&=&
\pm
\frac{\sqrt{3}r}
       {4\sqrt{6Mr+r^4 V}}
\left[
\frac{ (2C_0^3 c_1+3C_0^2c_2-c_4) (6M+3r+r^3V)r^3}    
         {(3r-6M-r^3V)^2}
+
2b_1(r)
\right],
\end{eqnarray} 
where $b_1(r)$ is a pure function of $r$, and 
the nontrivial components
of the Schwarzschild-de Sitter metrics
written in the nonstandard coordinates 
can be expanded in terms of $m^2$ as
\begin{eqnarray}
A_0(r)
&=&
A_{0f}(r)
=
1-b_1(r)m^2
+
{\cal O} (m^4),
\nonumber\\
A_1(r)
&=&
A_{1f}(r)
=
1+b_1(r)m^2
+
{\cal O} (m^4),
\nonumber\\
N^r(r)
&=&
\pm
\sqrt{\frac{2M}{r}+\frac{r^2 V}{3}}
\nonumber\\
&\pm &
m^2
\left[
\frac{\sqrt{3} \left(
  r^3\left(-2C_0^3c_1-3C_0^2 c_2+c_4\right)
-2 \left(6M+3r+r^3V\right)b_1(r)
\right)}
        {12\sqrt{r\left(6M +r^3V\right)}}
+
\frac{q_{n_{0}}(t)}{2}
\sqrt{\frac{2M}{r}+\frac{V}{3}r^2}
\right]
+
{\cal O} (m^4),
\nonumber\\
N_f^r(r)
&=&
\pm 
\sqrt{\frac{2M_f}{r}+\frac{r^2C_0^2V_f }{3}}
\nonumber\\
&\pm &
m^2
\left[
\frac{\sqrt{3}
\left(
r^3\left(C_0^2c_0+2C_0 c_1+c_2\right)
-2{\tilde\alpha}^2 (6M_f+3r+C_0^2r^3 V_f) b_{1}(r)
\right)}
        {12 {\tilde\alpha}^2 \sqrt{r\left(6M_f +r^3C_0^2V_f\right)}}
+
\frac{q_{n_{0f}}(t)}{2}
\sqrt{\frac{2M_f}{r}+\frac{C_0^2 V_{f}}{3}r^2}
\right]
\nonumber\\
&+&
{\cal O} (m^4),
\end{eqnarray}
which agrees with
Eq.\ \eqref{sol_sa_1}
with $q_{n^r}=q_{n^r_f}=0$ and $a_2(t,f)=a_{2f}(t,r)=0$.
Thus, 
the solution \eqref{mtbg_nonstandard}
is 
the nonperturbative extension of
the perturbative solution \eqref{sol_sa_1}
discussed in the previous section.

As an extension of our analysis in this section, 
we could start from even more general description of 
the Schwarzschild-de Sitter metrics 
in the physical and fiducial sectors
\begin{eqnarray}
\label{sch_ds_static2}
g_{\mu\nu}dx^\mu dx^\nu
&=&
-
\left(
1
-\frac{2M}{R(t,r)}
+\frac{V_{\rm eff}}{3}R(t,r)^2
\right)
dT(t,r)^2
+
\frac{dR(t,r)^2}
{1-\frac{2M}{R(t,r)}
+\frac{V_{\rm eff}}{3}R(t,r)^2}
+
R(t,r)^2
\left(
d\theta^2
+\sin^2\theta d\varphi^2
\right),
\nonumber\\
f_{\mu\nu}dx^\mu dx^\nu
&=&
C_0^2
\left[
-
\left(
1
-\frac{2M_f}{R(t,r)}
+\frac{C_0^2 V_{f,{\rm eff}}}{3}R_f(t,r)^2
\right)
dT_f(t,r)^2
\right.
\nonumber\\
&&
\left.
+
\frac{dR_f(t,r)^2}
{1-\frac{2M_f}{R_f(t,r)}
+\frac{C_0^2V_{f,{\rm eff}}}{3}R_f(t,r)^2}
+
R_f(t,r)^2
\left(
d\theta^2
+\sin^2\theta d\varphi^2
\right)
\right],
\end{eqnarray}
where $R(t,r)$ and $R_f(t,r)$ are the functions of $(t,r)$.
We expect that 
in the self-accelerating branch \eqref{additional},
the transformed metrics written 
in terms of $(t,r)$
satisfy 
all the metric and constraint equations
with the conditions \eqref{gf}, \eqref{veff}, and \eqref{lambda_conditions_three}.
Then,
$R(t,r)$ and $R_f(t,r)$
would correspond to  $a_2(t,r)$ and $a_{2f}(t,r)$
at ${\cal O} (m^2)$
in the perturbative approach, respectively.

\section{Conclusions}
\label{sec6}

We have investigated static and spherically symmetric solutions
in the Minimal Theory of Bigravity (MTBG).
Our main focus was
whether static and spherically symmetric solutions in 
general relativity (GR)
written in the standard or nonstandard coordinates, 
e.g., the spatially-flat Gullstrand-Painlev\'e (GP) coordinates,
can also be solutions in MTBG.
We have considered 
both the vacuum solutions and the solutions with matter.

First,
we have investigated 
the existence of vacuum Schwarzschild-de Sitter solutions
written in the spatially-flat GP coordinates in MTBG.
In the general static and spherically symmetric backgrounds
written in the spatially-flat coordinates \eqref{solution_both_constant1},
we have found that 
all the components of the constraint equation \eqref{const} were trivially satisfied.
In static and spherically symmetric backgrounds,
there could be two branches,
namely,
the self-accelerating and normal branches.
We have shown
that 
in the self-accelerating branch
the substitution of 
the algebraic conditions \eqref{cond} and \eqref{additional}
and 
the Schwarzschild-de Sitter metrics written in the spatially-flat coordinates
into the gravitational equations of motion in MTBG
results in a consistent solution, 
provided Eq.\ \eqref{all_lambda_vanish}. 
These results could be extended to the presence of matter in both sectors, thus allowing for static and spherically-symmetric stellar solutions in the self-accelerating branch of MTBG.
On the other hand,
in the normal branch
the Lagrange multiplier ${\bar \lambda}$ 
could not be zero when we chose $\lambda=0$,
and hence 
the Schwarzschild solutions 
written in the GP coordinates
could not satisfy the Euler-Lagrange equations in MTBG,
unless the masses of black holes (BHs) vanish. 

Since MTBG in the unitary gauge does not enjoy two copies of 
the four-dimensional diffeomorphism invariance, 
the absence of certain GR solutions
in a particular coordinate system
does not necessarily 
mean their absence in the other coordinates.
For instance, when written
in the Schwarzschild coordinates,
the Schwarzschild-de Sitter solutions
with equal ADM masses 
do satisfy 
the Euler-Lagrange equations in vacuum MTBG
under the condition \eqref{eq:defC}, even in the normal branch. A second example, this one regular at the horizon, is the Schwarzschild-de Sitter solution written in the slicing \eqref{VCDM_slicing} as in \cite{DeFelice:2020onz}, which is also a solution in the normal branch under the parallel metrics ansatz, i.e.\ $f_{\mu\nu}\propto g_{\mu\nu}$ (together with vanishing $\lambda$, $\partial_i\bar\lambda$ and $\lambda^i$). This normal-branch ansatz provides a solution not only for spherically symmetric configurations, but coincide with vacuum-GR solutions provided a slicing satisfying \eqref{VCDM_slicing} such as a constant mean curvature slicing is adopted. It remains to be investigated whether these solutions can be obtained via a standard matter collapse, and in such a sense are connected to more generic solutions. Aside from studying matter collapse directly, an avenue of research could be a perturbative detuning of the conditions \eqref{VCDM_slicing}. More in general, although we have provided a large class of physically interesting solutions (especially for the self-accelerating branch), still the most general solutions of MTBG in a spherically symmetric configuration are not known. For instance, non-perturbative solutions with $\lambda\neq0$ (and/or $\partial_i\bar\lambda\neq0$, and/or $\lambda^i\neq0$) are not known. Although a mathematical general solution seems hard to be found, what we can guess is that these solutions, if existing, would probably differ from vacuum GR solutions.

Finally, we have investigated whether MTBG 
admits nontrivial static and spherically symmetric solutions 
besides the GR solutions. 
We have constructed static and spherically symmetric vacuum solutions
perturbatively in the small mass limit of MTBG.
We have regarded the graviton mass squared as the expansion parameter,
and expanded the metric variables and the Lagrange multipliers.
We have shown that 
in the self-accelerating branch the nontrivial solutions are 
given by the Schwarzschild-de Sitter metrics written in nonstandard coordinates, 
while in the normal branch the spatially-flat Schwarzschild-de Sitter solutions are not compatible
with the set up of MTBG in the massless limit.
We have also confirmed
that 
in the self-accelerating branch 
the Schwarzschild-de Sitter solutions written 
in terms of the nonstandard coordinates 
with a single free function of $r$
satisfy all the metric and constraint equations in  MTBG
and 
correspond to 
the straightforward nonperturbative extension 
of the perturbative Schwarzschild-de Sitter solutions
in the small mass expansion.

Although we have obtained the Schwarzschild-de Sitter 
and
static spherically symmetric GR stellar solutions in MTBG,
the behavior of the linear perturbations 
against these solutions
could be different from that in GR.
A linear instability of a GR solution
may signal the realization of spontaneous tensorization,
which could provide a new way to probe
the existence of the graviton mass
in the strong gravity regime.
It would also be interesting 
to investigate the existence of GR solutions beyond 
the cosmological and the spherically symmetric configurations in MTBG,
for instance, stationary and axisymmetric solutions.
We hope that we will come back to these issues in our future work.

\begin{acknowledgments}
 The authors would like to thank Fran\c{c}ois Larrouturou for stimulating discussions,
 valuable comments, and proofreading the manuscript. 
 M.M.\ was supported by the Portuguese national fund through the
 Funda\c{c}\~{a}o para a Ci\^encia e a Tecnologia (FCT) in the scope of the
 framework of the Decree-Law 57/2016 of August 29, changed by Law
 57/2017 of July 19, and the Centro de Astrof\'{\i}sica e Gravita\c
 c\~ao ~(CENTRA) through the Project~No.~UIDB/00099/2020. 
M.M. also thanks Yukawa Institute for Theoretical Physics for the hospitality
 under the Visitors Program of FY2021.
 The work of A.D.F.\ was supported by Japan Society for the Promotion
 of Science Grants-in-Aid for Scientific Research No.~20K03969.  
 The work of S.M.~was supported in part by JSPS Grants-in-Aid for Scientific Research No.~17H02890, No.~17H06359, and by World Premier International Research Center Initiative, MEXT, Japan. 
\end{acknowledgments}

\appendix

\section{Non-existence of the spatially-flat Schwarzschild (-de Sitter) solutions in the massless limit of the normal branch of MTBG}
\label{app:nogo-normalbranch}

In this Appendix, we employ the small $m^2$ expansion in the normal branch of MTBG 
with $C_0^2c_1+2C_0c_2+c_3\neq 0$ and 
show the nonexistence of Schwarzschild BH solutions
written in the spatially-flat coordinates in the $m^2\to 0$ limit. 
While we mainly focus on the case 
where the background spacetimes at ${\cal O} (m^0)$
are given by the Schwarzschild metrics with  $V=V_f=0$
in Eq.\ \eqref{perturbative_expansion},
the extension to the Schwarzschild-de Sitter metrics with $V\neq 0$ and $V_f\neq 0$
is straightforward.

At ${\cal O}(m^2)$, 
the combinations  of the Euler-Lagrange equations
for 
$N^r$ and $N^r_f$
provide the general solution
\begin{eqnarray}
\label{sols_shift}
\Lambda_{-2}(t,r)
=\lambda_0(t),
\quad 
{\bar \Lambda}_{-2} (t,r)
=
d_1(t)
+
\frac{d_2(t)}{r}
+
d_3(t) r^2,
\end{eqnarray}
where
$\left(\lambda_0(t),d_1(t), d_2(t), d_3(t)\right)$
are pure functions of the time $t$.
At ${\cal O} (m^0)$,
the Euler-Lagrange equations for 
$A_0$ and $A_{0f}$
uniquely fix
\begin{eqnarray}
\label{sols_lapse}
\lambda_0(t)=0,
\quad 
d_3(t)=0.
\end{eqnarray}
At ${\cal O} (m^0)$,
a combination of the Euler-Lagrange equations
for $A_1$ and $A_{1f}$
reduces to 
\begin{eqnarray}
\left(
C_0^2c_1+2C_0 c_2+c_3
\right)
\left(
\sqrt{M_f}
+
C_0^2 \sqrt{M}
\right)
d_2(t)
=0,
\end{eqnarray}
which 
with $C_0^2c_1+2C_0 c_2+c_3\neq 0$
imposes
\begin{eqnarray}
\label{radial_sols}
d_2(t)=0.
\end{eqnarray}
Then, 
at ${\cal O} (m^0)$,
the Euler-Lagrange equations
for $A_1$ and $A_{1f}$ reduce to
\begin{eqnarray}
\label{bifurcation}
\left(
c_2+c_3\beta
+
C_0
\left(
c_1
+c_2\beta
\right)
\right)
{\Lambda}^r_{-2}
(t,r)
=0.
\end{eqnarray}
In Case \ref{case1},
because of $c_2+c_3\beta+
C_0
\left(
c_1
+c_2\beta
\right)\neq 0$,
we have to impose
\begin{eqnarray}
\label{lambdam2}
{\Lambda}^r_{-2}(t,r)=0.
\end{eqnarray}
At ${\cal O} (m^0)$,
the above equations
\eqref{sols_lapse},
\eqref{radial_sols},
and 
\eqref{lambdam2}
also satisfy the Euler-Lagrange equations for
$A_2$ and $A_{2f}$.
Thus,
at ${\cal O} (m^0)$,
all the metric Euler-Lagrange equations are satisfied.
Finally,
at ${\cal O} (m^2)$
the constraint equations \eqref{cons2} and \eqref{cons3}
provide  
\begin{eqnarray}
m^2
\left(
C_0^2c_1 +2C_0c_2+c_3
\right)
\left(
\sqrt{M}-\sqrt{M_f}
\right)
=0,
\quad
m^2
\left(
C_0^2c_1 +2C_0c_2+c_3
\right)
\left(
C_0^2\sqrt{M}+\sqrt{M_f}
\right)
=0,
\end{eqnarray}
which results in the no-go result
for the existence of the Schwarzschild metrics
written 
in the GP coordinates in the $m^2\to 0$ limit,
\begin{eqnarray}
\label{vanishing_mass}
M=M_f=0.
\end{eqnarray}
We note that 
adding the nonzero cosmological constants
$V\neq 0$ and $V_f\neq 0$ also provides the no-go result in the $m^2\to 0$ limit
\begin{eqnarray}
\label{vanishing_mass2}
M=M_f=0,
\quad 
\sqrt{V}=C_0\sqrt{V_f}.
\end{eqnarray}

In Case \ref{case2},
Eq.~\eqref{bifurcation} is automatically satisfied.
Then, 
at ${\cal O} (m^2)$,
the constraint equations \eqref{cons2} and \eqref{cons3}
provide
\begin{eqnarray}
m^2
\left(
\sqrt{M_f}-\sqrt{M}
\right)
\left(
C_0\beta-1
\right)=0,
\qquad
m^2
\left(
\sqrt{M_f}+C_0^2 \sqrt{M}
\right)
\left(
C_0\beta-1
\right)=0,
\end{eqnarray} 
which for $C_0\beta\neq 1$ 
again results in  
the no-go result
for the Schwarzschild solution \eqref{vanishing_mass}
written in the spatially flat coordinates.
Again, 
adding the nonzero cosmological constants
$V\neq 0$ and $V_f\neq 0$ also provides the no-go result
\eqref{vanishing_mass2} in the $m^2\to 0$ limit.

On the other hand, 
Case \ref{case3} with $C_0\beta=1$,
for which 
all the constraints \eqref{cons2}, \eqref{cons3} and \eqref{const}
are automatically satisfied,
corresponds to the special case
of the self-accelerating branch
and
needs not to be considered here separately.
Thus,
in the normal branch,
both Case \ref{case1} and Case \ref{case2}
result in the no-go result
that 
does not allow the existence of
Schwarzschild and Schwarzschild-de Sitter metrics
written in the spatially-flat coordinates
in the massless limit of MTBG.

\bibliography{refs}
\end{document}